\documentclass[amsmath,amssymb,nofootinbib,twocolumn,superscriptaddress,
  floatfix]{revtex4}
\usepackage{graphicx}% Include figure files
\usepackage{dcolumn}% Align table columns on decimal point
\usepackage{bm}% bold math
\usepackage{epsf}
\usepackage{rotating}
\usepackage{tabularx}
\usepackage{amsmath}
\usepackage{amssymb}
\usepackage[polish,american]{babel} % this is to get Zbyszek Was's name right...
\DeclareTextSymbolDefault{\SImu}{TS1}
\DeclareTextSymbol{\SImu}{TS1}{181} 
\DeclareFontEncoding{TS1}{}{}
\DeclareFontSubstitution{TS1}{cmr}{m}{n}
\DeclareFontFamily{OML}{eur}{\skewchar\font127} \DeclareFontShape{OML}{eur}{m}{n}{<5> <6>
                <7> <8> <9> gen * eurm <10><10.95><12><14.4><17.28><20.74><24.88>eurm10}{}
\DeclareSymbolFont{greek}{OML}{eur}{m}{n}
\renewcommand\eqref[1]{(\ref{#1})} % {Eq#2.\ \ref{#1}}
\newcommand\comment[1]{}
\def\micro{\SImu}
\newcommand\pt{\ensuremath{p_{T}}}
\newcommand\blind[1]{#1}

\newcommand\BR[1]{\ensuremath{{\cal B}(#1)}}%\left(#1\right)}}
\def\ed2{\ensuremath{\epsilon D^2}}

\def\Vud{V_{ud}}  \def\Vub{V_{ub}}
\def\Vcd{V_{cd}}  \def\Vcb{V_{cb}}
  
\def\CP{\ensuremath{C\!P}}

\def\Ds{\ensuremath{D_s^+}}
\def\Bs{\ensuremath{B^0_s}}
\def\Bsbar{\ensuremath{\overline{B}{}^0_s}}
\def\B0m{\ensuremath{B^0}}
\def\B0bar{\ensuremath{\overline{B}{}^0}}
\def\D0{\ensuremath{D^0}}
\def\D0bar{\ensuremath{\overline{D}{}^0}}

\def\photos{\textsc{photos}}
\def\phipiresult{\blind{0.097} \pm 0.018 ({\rm stat}) \pm 0.009 ({\rm sys})}
\def\bsdsk{\ensuremath{\Bsbar \to D_s^\pm K^\mp}}
\def\bsdspi{\ensuremath{\Bsbar \to D_s^+ \pi^-}}
\def\bsdsx{\ensuremath{\Bsbar \to D_s^+ X}}
\def\bsdsh{\ensuremath{\Bsbar \to D_s^\pm h^\mp}}

\def\bdpi{\ensuremath{\B0bar \to D^+ \pi^-}}
\def\bdx{\ensuremath{\B0bar \to D^+ X}}

\def\bdstarpi{\ensuremath{\B0bar \to D^{*+} \pi^-}}
\def\bdstarx{\ensuremath{\B0bar \to D^{*+} X}}

\renewcommand\eqref[2]{(\ref{#1})} % {Eq#2.\ \ref{#1}}

\newcommand\tableref[2]{Table#2~\ref{#1}}
\newcommand\figref[2]{Figure#2~\ref{#1}}

\begin{document}
\title{\bf First observation of {\boldmath $\Bsbar\rightarrow D_s^\pm K^\mp$}
  and measurement of the ratio of branching fractions
{\boldmath  \(\BR{\Bsbar\rightarrow D_s^\pm K^\mp} / 
\BR{\Bsbar\rightarrow D_s^{+}\pi^-}\)}}
\affiliation{Institute of Physics, Academia Sinica, Taipei, Taiwan 11529, Republic of China} 
\affiliation{Argonne National Laboratory, Argonne, Illinois 60439} 
\affiliation{University of Athens, 157 71 Athens, Greece} 
\affiliation{Institut de Fisica d'Altes Energies, Universitat Autonoma de Barcelona, E-08193, Bellaterra (Barcelona), Spain} 
\affiliation{Baylor University, Waco, Texas  76798} 
\affiliation{Istituto Nazionale di Fisica Nucleare Bologna, $^t$University of Bologna, I-40127 Bologna, Italy} 
\affiliation{Brandeis University, Waltham, Massachusetts 02254} 
\affiliation{University of California, Davis, Davis, California  95616} 
\affiliation{University of California, Los Angeles, Los Angeles, California  90024} 
\affiliation{University of California, San Diego, La Jolla, California  92093} 
\affiliation{University of California, Santa Barbara, Santa Barbara, California 93106} 
\affiliation{Instituto de Fisica de Cantabria, CSIC-University of Cantabria, 39005 Santander, Spain} 
\affiliation{Carnegie Mellon University, Pittsburgh, PA  15213} 
\affiliation{Enrico Fermi Institute, University of Chicago, Chicago, Illinois 60637} 
\affiliation{Comenius University, 842 48 Bratislava, Slovakia; Institute of Experimental Physics, 040 01 Kosice, Slovakia} 
\affiliation{Joint Institute for Nuclear Research, RU-141980 Dubna, Russia} 
\affiliation{Duke University, Durham, North Carolina  27708} 
\affiliation{Fermi National Accelerator Laboratory, Batavia, Illinois 60510} 
\affiliation{University of Florida, Gainesville, Florida  32611} 
\affiliation{Laboratori Nazionali di Frascati, Istituto Nazionale di Fisica Nucleare, I-00044 Frascati, Italy} 
\affiliation{University of Geneva, CH-1211 Geneva 4, Switzerland} 
\affiliation{Glasgow University, Glasgow G12 8QQ, United Kingdom} 
\affiliation{Harvard University, Cambridge, Massachusetts 02138} 
\affiliation{Division of High Energy Physics, Department of Physics, University of Helsinki and Helsinki Institute of Physics, FIN-00014, Helsinki, Finland} 
\affiliation{University of Illinois, Urbana, Illinois 61801} 
\affiliation{The Johns Hopkins University, Baltimore, Maryland 21218} 
\affiliation{Institut f\"{u}r Experimentelle Kernphysik, Universit\"{a}t Karlsruhe, 76128 Karlsruhe, Germany} 
\affiliation{Center for High Energy Physics: Kyungpook National University, Daegu 702-701, Korea; Seoul National University, Seoul 151-742, Korea; Sungkyunkwan University, Suwon 440-746, Korea; Korea Institute of Science and Technology Information, Daejeon, 305-806, Korea; Chonnam National University, Gwangju, 500-757, Korea} 
\affiliation{Ernest Orlando Lawrence Berkeley National Laboratory, Berkeley, California 94720} 
\affiliation{University of Liverpool, Liverpool L69 7ZE, United Kingdom} 
\affiliation{University College London, London WC1E 6BT, United Kingdom} 
\affiliation{Centro de Investigaciones Energeticas Medioambientales y Tecnologicas, E-28040 Madrid, Spain} 
\affiliation{Massachusetts Institute of Technology, Cambridge, Massachusetts  02139} 
\affiliation{Institute of Particle Physics: McGill University, Montr\'{e}al, Canada H3A~2T8; and University of Toronto, Toronto, Canada M5S~1A7} 
\affiliation{University of Michigan, Ann Arbor, Michigan 48109} 
\affiliation{Michigan State University, East Lansing, Michigan  48824}
\affiliation{Institution for Theoretical and Experimental Physics, ITEP, Moscow 117259, Russia} 
\affiliation{University of New Mexico, Albuquerque, New Mexico 87131} 
\affiliation{Northwestern University, Evanston, Illinois  60208} 
\affiliation{The Ohio State University, Columbus, Ohio  43210} 
\affiliation{Okayama University, Okayama 700-8530, Japan} 
\affiliation{Osaka City University, Osaka 588, Japan} 
\affiliation{University of Oxford, Oxford OX1 3RH, United Kingdom} 
\affiliation{Istituto Nazionale di Fisica Nucleare, Sezione di Padova-Trento, $^u$University of Padova, I-35131 Padova, Italy} 
\affiliation{LPNHE, Universite Pierre et Marie Curie/IN2P3-CNRS, UMR7585, Paris, F-75252 France} 
\affiliation{University of Pennsylvania, Philadelphia, Pennsylvania 19104} 
\affiliation{Istituto Nazionale di Fisica Nucleare Pisa, $^q$University of Pisa, $^r$University of Siena and $^s$Scuola Normale Superiore, I-56127 Pisa, Italy} 
\affiliation{University of Pittsburgh, Pittsburgh, Pennsylvania 15260} 
\affiliation{Purdue University, West Lafayette, Indiana 47907} 
\affiliation{University of Rochester, Rochester, New York 14627} 
\affiliation{The Rockefeller University, New York, New York 10021} 

\affiliation{Istituto Nazionale di Fisica Nucleare, Sezione di Roma 1, $^v$Sapienza Universit\`{a} di Roma, I-00185 Roma, Italy} 

\affiliation{Rutgers University, Piscataway, New Jersey 08855} 
\affiliation{Texas A\&M University, College Station, Texas 77843} 
\affiliation{Istituto Nazionale di Fisica Nucleare Trieste/\ Udine, $^w$University of Trieste/\ Udine, Italy} 
\affiliation{University of Tsukuba, Tsukuba, Ibaraki 305, Japan} 
\affiliation{Tufts University, Medford, Massachusetts 02155} 
\affiliation{Waseda University, Tokyo 169, Japan} 
\affiliation{Wayne State University, Detroit, Michigan  48201} 
\affiliation{University of Wisconsin, Madison, Wisconsin 53706} 
\affiliation{Yale University, New Haven, Connecticut 06520} 
\author{T.~Aaltonen}
\affiliation{Division of High Energy Physics, Department of Physics, University of Helsinki and Helsinki Institute of Physics, FIN-00014, Helsinki, Finland}
\author{J.~Adelman}
\affiliation{Enrico Fermi Institute, University of Chicago, Chicago, Illinois 60637}
\author{T.~Akimoto}
\affiliation{University of Tsukuba, Tsukuba, Ibaraki 305, Japan}
\author{M.G.~Albrow}
\affiliation{Fermi National Accelerator Laboratory, Batavia, Illinois 60510}
\author{B.~\'{A}lvarez~Gonz\'{a}lez}
\affiliation{Instituto de Fisica de Cantabria, CSIC-University of Cantabria, 39005 Santander, Spain}
\author{S.~Amerio$^u$}
\affiliation{Istituto Nazionale di Fisica Nucleare, Sezione di Padova-Trento, $^u$University of Padova, I-35131 Padova, Italy} 

\author{D.~Amidei}
\affiliation{University of Michigan, Ann Arbor, Michigan 48109}
\author{A.~Anastassov}
\affiliation{Northwestern University, Evanston, Illinois  60208}
\author{A.~Annovi}
\affiliation{Laboratori Nazionali di Frascati, Istituto Nazionale di Fisica Nucleare, I-00044 Frascati, Italy}
\author{J.~Antos}
\affiliation{Comenius University, 842 48 Bratislava, Slovakia; Institute of Experimental Physics, 040 01 Kosice, Slovakia}
% ...

\collaboration{CDF Collaboration\footnote{With visitors from $^a$Universiteit Antwerpen, B-2610 Antwerp, Belgium, 
$^b$Chinese Academy of Sciences, Beijing 100864, China, 
$^c$University of Bristol, Bristol BS8 1TL, United Kingdom, 
$^d$University of California Irvine, Irvine, CA  92697, 
$^e$University of California Santa Cruz, Santa Cruz, CA  95064, 
$^f$Cornell University, Ithaca, NY  14853, 
$^g$University of Cyprus, Nicosia CY-1678, Cyprus, 
$^h$University College Dublin, Dublin 4, Ireland, 
$^i$University of Edinburgh, Edinburgh EH9 3JZ, United Kingdom, 
$^j$Universidad Iberoamericana, Mexico D.F., Mexico, 
$^k$University of Manchester, Manchester M13 9PL, England, 
$^l$Nagasaki Institute of Applied Science, Nagasaki, Japan, 
$^m$University de Oviedo, E-33007 Oviedo, Spain, 
$^n$Queen Mary, University of London, London, E1 4NS, England, 
$^o$Texas Tech University, Lubbock, TX  79409, 
$^p$IFIC(CSIC-Universitat de Valencia), 46071 Valencia, Spain,
$^x$Royal Society of Edinburgh/Scottish Executive Support Research Fellow,
$^y$Istituto Nazionale di Fisica Nucleare, Sezione di Cagliari, 09042 Monserrato (Cagliari), Italy 
}}
\noaffiliation

\date{\today}
%% \maketitle
%
%
\begin{abstract}
A combined mass and particle identification fit is used to make the first
observation of the decay $\Bsbar\rightarrow D_s^\pm K^\mp$ and measure the
branching fraction of $\Bsbar \to D_s^\pm K^\mp$ relative to $\Bsbar \to
D_s^+\pi^-$.  This analysis uses 1.2 fb$^{-1}$ integrated luminosity of
$p\overline p$ collisions at 
$\sqrt s = 1.96$ TeV collected with the CDF II detector at
the Fermilab Tevatron collider.  We observe a \(\Bsbar\rightarrow D_s^\pm
K^\mp\) signal with a statistical significance of $8.1\sigma$ and measure
\(\BR{\Bsbar\rightarrow D_s^\pm K^\mp} / \BR{\Bsbar\rightarrow D_s^{+}\pi^-} =
\phipiresult \). 
\end{abstract}
%% pacs: specific reactions and phenomenology (bottom mesons),
%%       particle properties (bottom mesons),
%%       determination of CKM matrix elements
\maketitle 
\par One of the remaining open questions in flavor physics is the precise value
of the angle $\gamma = \arg\left(-\Vud\Vub^*/\Vcd\Vcb^*\right)$ of the unitarity
triangle.  Current measurements use the interference between $b\to u\bar{c}s$
and $b\to c\bar{u}s$ diagrams in $B^- \to D^{(*)0} K^{(*)-}$ and $B^- \to
\overline{D}{}^{(*)0} K^{(*)-}$ decays when $D^0$ and \D0bar\ are observed in
common final states \cite{Gronau:1990ra, Gronau:1991dp, Atwood:1996ci,
  Giri:2003ty, Charles:2004jd, Bona:2006ah}, but suffer from the large
difference between the amplitudes of these decays.  With a large sample of
hadronic $\Bsbar$ decays, it may be possible to determine $\gamma$ from the
interference, through \Bs--\Bsbar\ mixing, of the same diagrams in the decay
modes $\Bsbar\to D_s^+ K^-$ and $\Bsbar\to D_s^- K^+$ \cite{Aleksan:1991nh,
  Dunietz:1995cp}, which are expected to have a more favorable amplitude ratio;
the two decays proceed through color-allowed tree amplitudes whose ratio is
suppressed by only a factor $\sim 0.4$ \cite{Fleischer:2003yb}.  To determine
$\gamma$, a time-dependent measurement of the decay rates of $\Bsbar \to D_s^+
K^-$, $\Bsbar\to D_s^- K^+$, $\Bs\to D_s^-K^+$, and $\Bs\to D_s^+ K^-$ is
required.  The first steps in this effort are to observe these decay modes
(which we will collectively refer to as \bsdsk) and to measure the \CP-averaged
branching ratio \(\BR{\bsdsk} \equiv \frac12[\BR{\Bsbar \to D_s^+ K^-} +
  \BR{\Bsbar\to D_s^- K^+} + \BR{\Bs\to D_s^-K^+} + \BR{\Bs\to D_s^+ K^-}]\).
In this Letter we report the first observation of the \bsdsk\ decay modes and
the first measurement of \(\BR{\bsdsk} / \BR{\bsdspi}\).  We measure this
branching fraction ratio since many of the systematic uncertainties cancel in
the ratio and $\BR{\bsdspi}$ is precisely measured elsewhere
\cite{Abulencia:2006qw, Drutskoy:2006xc}.
%
%%
%%
%%%%%%%%%%%%%%%%%%%%%%%%%%%%%%%%%%%%%%%%%%%%%%%%%%%%%%%%%%%%%%%%%%%%%%
%%%%%%%%%%%%%%%%%%%%%%%%%%%%%%%%%%%%%%%%%%%%%%%%%%%%%%%%%%%%%%%%%%%%%%
%\section{Data sample and event selection}\label{sec:data}
%
\par We analyze $p\overline p$ collisions at $\sqrt s = 1.96$ TeV recorded by
the CDF II detector at the Fermilab Tevatron collider with an integrated
luminosity of 1.2~fb$^{-1}$.  A detailed description of the detector can be
found elsewhere \cite{Acosta:2004yw}.  This analysis uses charged particle
tracks reconstructed in the pseudorapidity \cite{note:coords} range $|\eta|
\lesssim 1$ from hits in a silicon microstrip vertex detector \cite{Sill:2000zz}
and a cylindrical drift chamber \cite{Affolder:2003ep} immersed in a 1.4 T axial
magnetic field.  The specific ionization energy loss ($dE/dx$) of charged
particles in the drift chamber is used for particle identification (PID).  A
sample rich in bottom hadrons is selected by triggering on events that have at
least two tracks, each with $\pt >2$ GeV$/c$ and large impact parameter; the
trigger further requires that these tracks originate from a secondary vertex
well displaced from the primary interaction point \cite{Ashmanskas:2003gf}.
\par We reconstruct \bsdsh\ candidates (where $h=\pi$ or $K$) as follows.
First, we identify $D_s^+\to \phi(\to K^-K^+) \pi^+$ candidates
\cite{note:charge-conjugate} using the invariant mass requirements $1013 <
m(K^-K^+) < 1028\mbox{ MeV$/c^2$}$ and $1948.3 < m(K^-K^+\pi^+) < 1988.3\mbox{
  MeV$/c^2$}$.  The \Ds\ decay tracks must satisfy a three-dimensional vertex
fit.  No PID requirements are made on the \Ds\ decay tracks.  We then pair the
\Ds\ candidates with $h^-$ tracks to define the \bsdsh\ candidate sample, and
require the \Ds--$h^-$ pair to satisfy a three-dimensional fit to the
\Bsbar\ decay vertex.  No mass constraint is used either on the $\phi$ or on the
\Ds\ candidate.  Finally, we define a mass variable $m(D_s\pi)$ for the $D_s
\pi$ hypothesis (i.e., assigning the daughter track $h$ as a pion); $m(D_s\pi)$
is used to provide kinematic separation between the \bsdsk\ and
\bsdspi\ signals.
%% \bsdsh\ candidates (where $h = \pi$ or $K$) are reconstructed in the
%% $\phipi+\pi^-$ hypothesis, with $\phi\to K^+K^-$ \cite{note:charge-conjugate}.
%% Tracks are assumed to be kaons or pions to match the reconstruction hypothesis.
%% No explicit particle identification is used in the reconstruction and selection
%% of $\Bsbar$ candidates.  Tracks are combined to form $\phi$, $D_s^+$, and
%% $\Bsbar$ candidates.  The track combinations must be consistent with forming
%% three-dimensional $D_s^+$ and $\Bsbar$ vertices.  Combinations falling outside a
%% window around the invariant mass of the $\phi$ or $D_s^+$ ($1013 < m(K^+K^-) <
%% 1028\mbox{ MeV$/c^2$}$ and $1948.3 < m(K^+K^-\pi^+) < 1988.3\mbox{ MeV$/c^2$}$)
%% are rejected.
%
\par Further selection requirements are made to reduce combinatorial background.  The
discriminating variables are the transverse ($|d_0| < 60$ \micro m) and
longitudinal ($|z_0/\sigma_{z_0}| < 3$, where $\sigma_{z_0}$ is the uncertainty
on $z_0$) impact parameter of the $\Bsbar$ candidate
with respect to the primary event vertex; the transverse momentum ($\pt > 5.5$
GeV$/c$) of the $\Bsbar$ candidate; the isolation of the $\Bsbar$ candidate
\[
I = \frac{\pt\left(\Bsbar\right)}{\pt\left(\Bsbar\right) + \sum\limits_{\rm tracks}
  \pt(\mbox{track})} > 0.5,
\]
where the sum runs over tracks within $\Delta R = \sqrt{\Delta \phi^2 + \Delta
  \eta^2} < 1$ around the $\Bsbar$ direction originating from the same primary
vertex; the opening angle [$\Delta R(D_s^+, h^-) < 1.5$] between the $D_s^+$
candidate and the track originating from the $\Bsbar$ decay vertex (the
``$\Bsbar$ daughter track''); and the projection of the $\Bsbar$ and $D_s^+$
decay length along the transverse momentum of the $\Bsbar$ candidate [$L_{xy}
  \left(\Bsbar\right) > 300$ \micro m, $L_{xy}
  \left(\Bsbar\right)/\sigma_{L_{xy}} \left(\Bsbar\right) > 8$ (where
  $\sigma_{L_{xy}}$ is the uncertainty on $L_{xy}$), and $L_{xy} (D_s^+) >
  L_{xy} \left(\Bsbar\right)$].  The $dE/dx$ calibrations are based on large
samples of $D^0\to K^-\pi^+$ decays taken with the displaced track trigger.  To
avoid bias, the \Bsbar\ daughter tracks are required to pass the same $\pt > 2$
GeV$/c$ trigger requirement as the $D^0\to K^-\pi^+$ calibration tracks.
%% Accurate $dE/dx$ calibrations are available only for
%% tracks that were also reconstructed by the trigger-level track finding; we
%% therefore require the $\Bsbar$ daughter tracks to be matched in $\phi$ and
%% \pt\ to trigger-level tracks.
%
\par Monte Carlo simulation is used to model signal and background and to
determine trigger and reconstruction efficiencies.  We generate single $\Bsbar$
hadrons with \textsc{bgenerator} \cite{Nason:1987xz, Nason:1989zy} and simulate
their decays with \textsc{evtgen} \cite{Lange:2001uf}.  A detailed detector and
trigger simulation is then performed.
%% Additional corrections to the Monte Carlo resolution and mass scale are
%% applied.  These corrections are determined by comparing the peak mean and
%% width in a clean \bdstarpi\ data sample with Monte Carlo.  For each sample
%% used in this analysis, the corrections are cross-checked \textit{in situ} by
%% scanning the maximum likelihood as a function of resolution and mass scale.
%%
%%
%%%%%%%%%%%%%%%%%%%%%%%%%%%%%%%%%%%%%%%%%%%%%%%%%%%%%%%%%%%%%%%%%%%%%%
%%%%%%%%%%%%%%%%%%%%%%%%%%%%%%%%%%%%%%%%%%%%%%%%%%%%%%%%%%%%%%%%%%%%%%
%\section{Analysis method}\label{sec:method}
%
\par The greatest challenge in this analysis is to disentangle the various
components contributing to the data sample.  Apart from the \bsdsk\ and
\bsdspi\ signals, the sample contains partially reconstructed \Bsbar\ decays,
reflections from decays of other bottom hadron species, and combinatorial
background.  To separate the components and determine the number of candidates
of each component type, we perform a maximum-likelihood fit.  The fit variables
are the invariant mass $m(D_s\pi)$ of the candidate in the $D_s\pi$ mass
hypothesis and the PID variable $Z$, which is the logarithm of the ratio between
the measured $dE/dx$ and the expected $dE/dx$ for a pion with the momentum of
the \Bsbar\ daughter track.  The likelihood function is
\begin{equation}
  L(f_1, \dots, f_{M - 1}) = \prod\limits_{i=1}^{N}\sum\limits_{j=1}^{M}  
  f_j\,p_j(m_i)\,q_j(Z_i), 
\label{eq-L}
\end{equation}
where $f_M = 1 - \sum_{j=1}^{M - 1} f_j$.  The index $i$ runs over the $N$
candidates, and $j$ runs over the $M$ components; $f_j$ is the fraction
of candidates in the $j$th component, to be determined by the fit.  
\par We group $\Bsbar$ candidates into three categories by source.
Combinations where the $D_s^+$ candidate and the track come from a single bottom
hadron ($\B0bar$, $B^-$, $\Bsbar$, $\Lambda_b^0$) are called single-$B$
contributions.  Non-bottom contributions where the $D_s^+$ candidate does not
come from a real $D_s^+$ are called fake-$D_s^+$ combinatorial background.
Combinations of a real $D_s^+$ with  a track
coming from fragmentation, the underlying event, or the other bottom hadron in
the event are called real-$D_s^+$ combinatorial background.
%
%% In the high-statistics, high-background control sample $\B0bar\to D^+X$, a
%% second combinatorial-background model (real-$D$ plus track taken from the
%% wrong-sign data sample) needed to be considered.  Although we know that a
%% combinatorial background model comprising two sources of background (fake $D$ +
%% track, real $D$ + track) is a better model than a single-source background, we
%% are limited in the $D_s\to\phi\pi$ case by the wrong-sign ($D_s^-\pi^-$) sample
%% statistics.  We find that the fit quality is adequate with a single, fake-$D_s$
%% background component.  In \secref{sec-fitdsphipi-syst}{} we assess a systematic
%% uncertainty due to the absence of a real-$D$ combinatorial background fit
%% component. 
%%
%%
%
%
%\subsection{Mass templates}\label{sec:method:masstemp}
\par Mass probability density functions (pdf's) $p_j(m)$ for the single-$\Bsbar$
components are extracted from large simulated samples of $\bsdsx$ decays, where
the $D_s^+$ is forced to decay to $\phi\pi^+$.  Separate mass templates are
extracted for \bsdspi\ and \bsdsk\ fully reconstructed decays and for the
partially reconstructed modes that overlap in mass with the \bsdsk: $\Bsbar\to
D_s^+\rho^-$ and $\Bsbar \to D_s^{*+} \pi^-$.  Partially reconstructed modes
missing more than one decay product are collected in one template.
Contributions from the $\B0bar\to D^+(K^-\pi^+\pi^+)X$, $B^-\to
D^+(K^-\pi^+\pi^+)X$ and $\Lambda_b^0 \to \Lambda_c^+ (p K^- \pi^+)X$
reflections are included.  Likewise, we include $\B0bar \to D_s^{(*)\pm}h^\mp$
decays (where $h=\pi,K$) whose branching fractions are known \cite{Aubert:2006hu};
the yields of these \B0bar\ modes relative to each other are fixed to the values
reported in \cite{Aubert:2006hu}.  Rather than parameterizing the mass shapes,
which are complicated for most of the single-$B$ components, we use histograms
as mass pdf's.  
Sufficiently large Monte Carlo samples [approximately 50\,000
  candidates after cuts of \bsdspi\ and \bsdsk] are generated to make the
statistical fluctuations in the pdf's small. 
\par Special care has to be taken in the treatment of the low-mass tail of the
decay mode $\bsdspi(n\gamma)$, which is dominated by the radiative tail, and
which overlaps with the \bsdsk\ mass pdf.
Improper accounting of the tail can bias both the measurement of the
\bsdspi\ yield and the \bsdsk\ yield by misidentifying a fraction of the
\bsdspi\ contribution as part of the (much smaller) \bsdsk\ contribution.  The
radiative tail is modeled in \textsc{evtgen} by using the \photos\ algorithm for
radiative corrections \cite{Barberio:1993qi} with a cut-off for photon emission
at 10 MeV.  We allow the normalization to float in the fit to account for
uncertainties in the \photos\ prediction of the size of the radiative tail.
(The radiative tail of \bsdsk\ does not require special treatment.  The kaon
radiates less, and any resulting misidentified \bsdsk\ contribution is easily
absorbed by the other fit components, which dominate at $m(D_s^+\pi^-)$ below
the \bsdsk\ peak.)
%
% \par Smearing, scale factor (determined from data, crosschecked by fit).
%
\par The mass distribution of the fake-$D_s^+$ background is parameterized with
a function of the form $p_{\rm bg}(m) \propto \exp(-\alpha m) + \beta$.  The
shape parameters $\alpha$ and $\beta$ are determined in an ancillary mass-only
fit of $\Bsbar$ candidates populating the sidebands of the $D_s^+$ mass
distribution.  To model the real-$D_s^+$ background, we use a sample of
same-sign $D_s^+\pi^+$ candidates.  A fit analogous to the fake-$D_s^+$ case is
performed on the $D_s^+\pi^+$ mass distribution.  Given to the limited
statistics of the signal sample, we cannot separately resolve the real-$D_s^+$
and fake-$D_s^+$ combinatorial backgrounds; in the default fit we therefore
combine the two types of background.  We assess a systematic uncertainty by
allowing the relative size of the two background types to vary.
%% Given the limited statistics, our analysis is not sensitive to a separate
%% real-$D_s^+$ combinatorial background component. In the default fit we set this
%% component to zero and assess a systematic uncertainty by treating its size as an
%% additional parameter of the fit.
%
\par
\begin{figure}[htp]
\centering\epsfxsize=\columnwidth\epsfbox{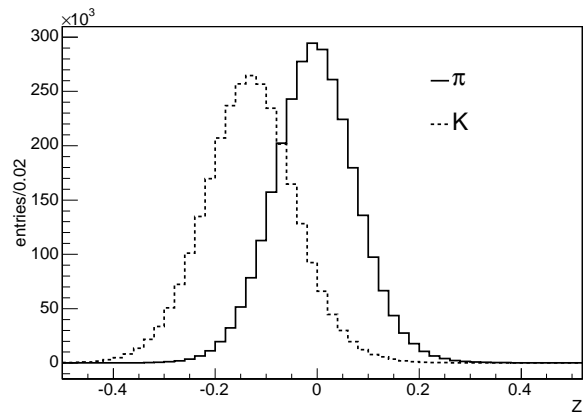}
\caption{$Z$ distributions for pions and kaons derived from prompt $D^{*+}$
  decays.}\label{fig:Zpdf} 
\end{figure}

 We determine the $Z$ pdf's $q_j(Z)$ for pions and kaons from $D^{*+}\to
D^0(K^-\pi^+)\pi^+$ decays.  The flavor of the daughter tracks of the $D^0$ in
the decay $D^{*+}\to D^0(K^-\pi^+)\pi^+$ is tagged by the charge of the soft
pion from the $D^{*+}$ decay.  Taken together with the large
signal-to-background ratio of the $\Delta m = m(K^-\pi^+\pi^+) - m(K^-\pi^+)$
peak, this charge tagging yields a very clean sample of pions and kaons.  We
further reduce background contamination by sideband-subtracting in $\Delta m$.
The mean values of $Z$ for kaons and pions are separated by approximately 1.4
standard deviations.  The $Z$ distributions for both species (shown in
\figref{fig:Zpdf}{}) have similar widths.
%% Hit-level and track-level corrections are applied to $Z$.  Two additional
%% effects are accounted for: First, the $Z$ of kaon tracks is
%% momentum-dependent, since the expected $dE/dx$ is calculated under an
%% incorrect mass hypothesis.  Second, differences in the tracking chamber
%% activity in the vicinity of the track of interest cause a small discrepancy
%% between tracks in the calibration $D^{*+}$ sample and in the physics $B\to
%% DX$ samples; this discrepancy is reduced by correcting for the track
%% ``secance'', or number of $r$--$\phi$ intersections with other tracks between
%% the COT inner and outer radii.
Because the data sample contains semileptonic decays, we need to
model the $dE/dx$ distributions of muons and electrons as well.  For muons,
which are a small contribution in the mass region of interest, the pion template
can be used without introducing a significant systematic uncertainty.  For
electrons, we derive a template from a $J/\psi\to e^+e^-$ sample.  The $Z$ pdf
for the fake-$D_s^+$ combinatorial background is determined from data by
selecting candidates from the sidebands of the $D_s^+$ mass distribution.  All
$Z$ pdf's are represented as histograms.
%
%\subsection{Efficiency corrections}\label{sec:method:corr}
%
\par
\begin{figure}[htp]
\centering\epsfxsize=\columnwidth\epsfbox{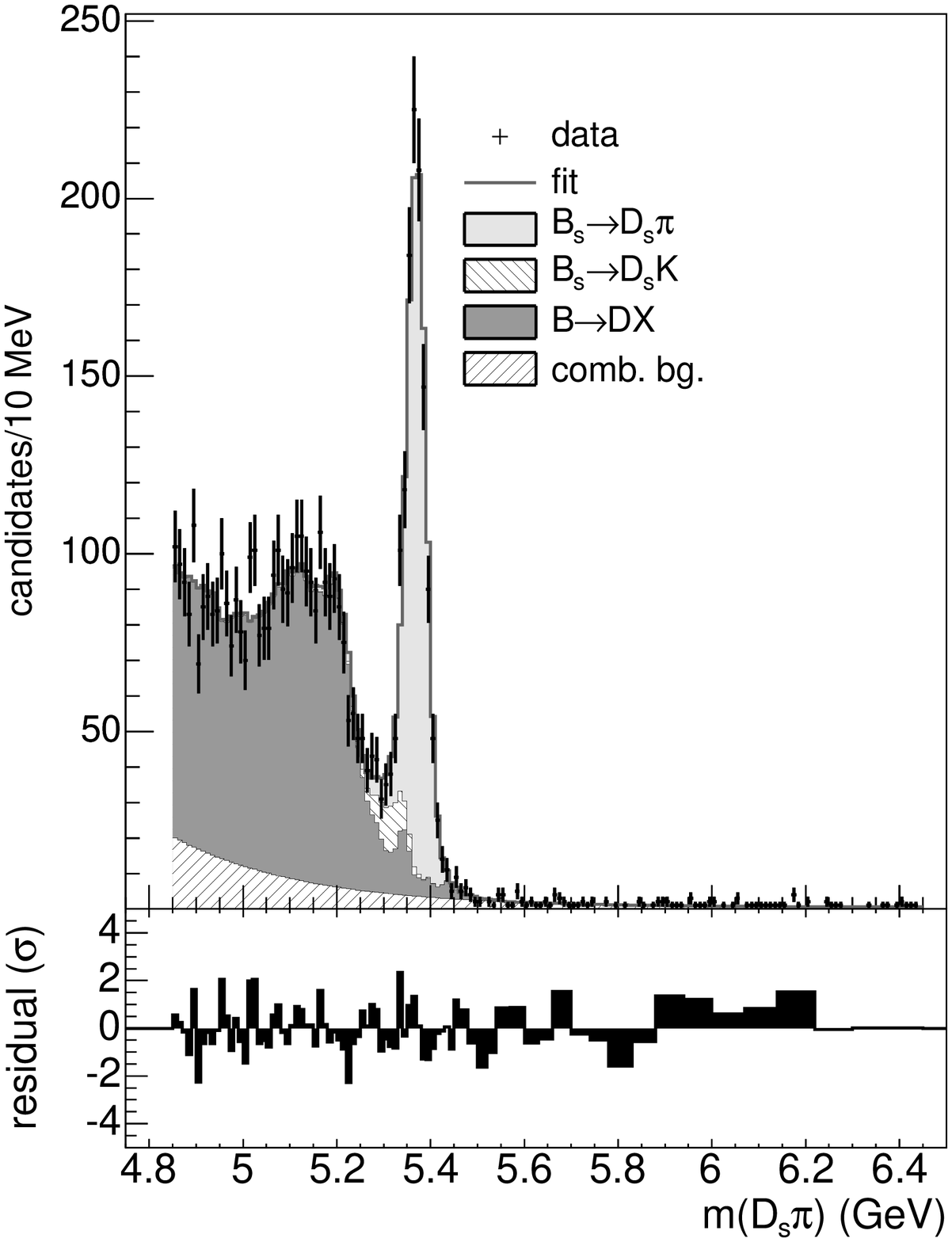}
\caption{Mass projection of the likelihood fit.  Fit components are stacked.  $B\to
  DX$ denotes all single-$B$ contributions except \bsdspi\ and \bsdsk, namely
  $\Bsbar\to D_s^+\rho^-$, $\Bsbar \to D_s^{*+} \pi^-$, partially reconstructed
  $\Bsbar \to D_s^+X$ modes missing more than one decay product, $\B0bar\to
  D^+(K^-\pi^+\pi^+)X$, $B^-\to D^+(K^-\pi^+\pi^+)X$ and $\Lambda_b^0 \to
  \Lambda_c^+ (p K^- \pi^+)X$ reflections, and $B^0 \to D_s^{(*)+}\pi^-$ and
  $B^0 \to D_s^{(*)-}K^+$; the small peak in the $B\to DX$ template is due to
  the $\B0bar\to D^+(K^-\pi^+\pi^+)\pi^-$ reflection.  The residual plot at the
  bottom shows the discrepancy (data minus fit) in units of standard deviation
  ($\sigma$); for the bins with low statistics, neighboring bins are combined
  until the predicted number of events is greater than five.  The $\chi^2$ of
  the projection is 79.0 for 72 degrees of freedom.}\label{fig:mass}
\end{figure}

\begin{figure}[htp]
\centering\epsfxsize=\columnwidth\epsfbox{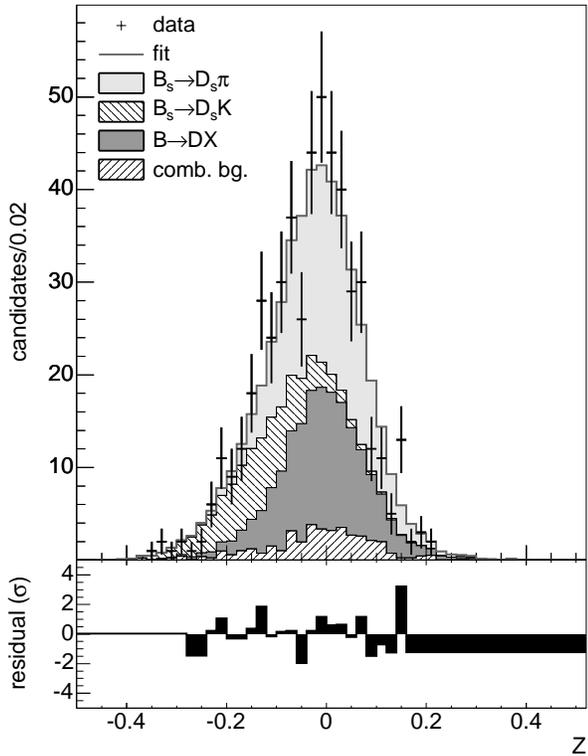}  
\caption{$Z$ projection of the likelihood fit in the region of interest for
  \bsdsk\ ($5.26 < m(D_s\pi) < 5.35\mbox{ GeV$/c^2$}$).  Fit
  components are stacked.  The residual plot at the bottom shows the discrepancy
  (data minus fit) in units of standard deviation ($\sigma$); for the bins with
  low statistics, neighboring bins are combined until the predicted number of
  events is greater than five.  The $\chi^2$ of the projection is 30.7 for 14
  degrees of freedom.}\label{fig:Z} 
\end{figure}

 \figref{fig:mass}{s} and \ref{fig:Z} show the fit projections in mass and
$Z$.  The yields determined by the fit are 
%% $14070\pm 189$ \bdpi, $1230\pm 58$
%% \bdk, $3728\pm 62$ \bdstarpi, $304\pm 25$ \bdstark, 
$1125\pm 87$ \bsdspi\ and $102\pm 18$ \bsdsk\ candidates.  The branching
fraction of \bsdsk\ relative to \bsdspi, corrected for the relative
reconstruction efficiency $\epsilon_\pi / \epsilon_K = 1.071\pm 0.028$, is
\(\BR{\Bsbar\rightarrow D_s^\pm K^\mp} / \BR{\Bsbar\rightarrow D_s^{+}\pi^-} =
0.097 \pm 0.018\).  A fit performed with the \bsdsk\ yield set to zero is worse
than the default fit by $\Delta \log L = -32.52$; the corresponding statistical
significance of the $\Bsbar\rightarrow D_s^\pm K^\mp$ signal is $8.1$ standard
deviations. 
\par Systematic uncertainties on $\BR{\bsdsk}/\BR{\bsdspi}$ are studied by
incorporating each effect in the generation of simulated experiments
%%  with mass
%% and $Z$ distributed according to alternate models, that 
which are then fitted using the default configuration.  The bias on
$\BR{\bsdsk}/\BR{\bsdspi}$, averaged over 10000 simulated experiments, is used
as the systematic uncertainty associated with the effect under study.
\tableref{tbl:syst}{} summarizes the systematic uncertainties.  The systematic
uncertainties are dominated by the modeling of the $dE/dx$ (0.007), specifically
by the differences between the $Z$ distributions of $D^{*+}$ daughter tracks
(from which the kaon and pion $Z$ pdf's are derived) and $\Bsbar$ daughter
tracks; these differences arise from effects such as the greater particle
abundance in the vicinity of a prompt $D^{*+}$ compared to a $\Bsbar$, and hence
a higher probability for $D^{*+}$ daughter tracks to contain extraneous hits.
Modeling of the mass distributions of the single-$B$ components (0.004), which
includes statistical fluctuations in the mass pdf's, and modeling of the
combinatorial-background mass shape (0.002) are comparatively minor
contributions.  The total systematic uncertainty is obtained by adding the
individual systematic uncertainties in quadrature; at 0.009, it is about half as
large as the statistical uncertainty.
\begin{table}
\caption{Systematic uncertainties on $\BR{\Bsbar\rightarrow
D_s^\pm K^\mp}/\BR{\Bsbar\rightarrow D_s^+\pi^-}$.}
\label{tbl:syst}
  \begin{center}
    \begin{tabular}{lc}
      \hline\hline
      Source	& Systematic uncertainty\\
      \hline
      $dE/dx$ pdf modeling				&	0.007	\\
      Mass pdf modeling			 	&	0.004	\\
      Combinatorial-background model			&	0.002	\\
      Fitter bias due to finite statistics		&	0.001	\\
      \hline						 	
      Sum in quadrature					& 	0.009   \\ 
      \hline\hline
    \end{tabular}
  \end{center}
\end{table}
\par The analysis procedure was crosschecked in several ways.  Most importantly,
before performing the measurement on the \bsdsk\ signal sample, we verified our
method using two control samples, \bdx\ and \bdstarx.   In both cases, our
results are statistically consistent with 
world-average values.  We measure \(\BR{\B0bar\rightarrow D^{+}K^-} /
\BR{\B0bar\rightarrow D^{+}\pi^-} = 0.086 \pm 0.005 ({\rm stat})\), 1.0 standard
deviations from the world average; and \(\BR{\B0bar\rightarrow D^{*+}K^-} /
\BR{\B0bar \rightarrow D^{*+}\pi^-} = 0.080 \pm 0.008 ({\rm stat})\), 0.3
standard deviations from the world average \cite{Amsler:2008zz}.
%% (These control-sample results are provided only as a cross-check of our
%% method.)
The relative branching fractions $\BR{\B0bar\to D^+\rho^-}/\BR{\bdpi}$,
$\BR{\B0bar\to D^{*+}\pi^-}/\BR{\bdpi}$, and $\BR{\B0bar\to
  D^{*+}\rho^-}/\BR{\bdstarpi}$ were also found to be consistent with world
averages.  Finally, the fractional size of the radiative tails of $\bdpi$,
$\bdstarpi$, and $\bsdspi$ are found to be in agreement with each
other (and about twice as large as the \photos\ prediction).
\par In conclusion, we have presented the first observation of the \bsdsk\ decay
mode with a statistical significance of $8.1$ standard deviations.  The
\bsdsk\ event yield is $102\pm18$ (statistical uncertainty only).  We use this
sample to measure \(\BR{\Bsbar\rightarrow D_s^\pm K^\mp} / \BR{\Bsbar\rightarrow
  D_s^{+}\pi^-} = \phipiresult \).  This result is consistent with naive
expectations based on the branching fraction ratio for the analogous $\B0bar$
and $B^-$ decays, taking into account also the expected contribution from
$\Bsbar \to D_s^-K^+$ decays \cite{Fleischer:2003yb}.
\begin{acknowledgments}
\par We thank the Fermilab staff and the technical staffs of the participating
institutions for their vital contributions. This work was supported by the
U.S.\ Department of Energy and National Science Foundation; the Italian Istituto
Nazionale di Fisica Nucleare; the Ministry of Education, Culture, Sports,
Science and Technology of Japan; the Natural Sciences and Engineering Research
Council of Canada; the National Science Council of the Republic of China; the
Swiss National Science Foundation; the A.P.\ Sloan Foundation; the
Bundesministerium f\"ur Bildung und Forschung, Germany; the Korean Science and
Engineering Foundation and the Korean Research Foundation; the Science and
Technology Facilities Council and the Royal Society, UK; the Institut National
de Physique Nucleaire et Physique des Particules/CNRS; the Russian Foundation
for Basic Research; the Ministerio de Ciencia e Innovaci\'{o}n, Spain; the
Slovak R\&D Agency; and the Academy of Finland.
\end{acknowledgments}
%%
%%
%% \bibliography{bsdsk}

\begin{thebibliography}{10}
\bibitem{note:coords}
CDF II uses a right-handed coordinate system with the origin at the center of
  the detector, in which the $z$ axis is along the proton direction, the $y$
  axis points up, $\theta$ and $\phi$ are the polar and azimuthal angles, and
  $r$ is the radial distance in the $x$--$y$ plane. The pseudorapidity $\eta$
  is defined as $-\log \tan (\theta/2)$.
\bibitem{note:charge-conjugate}
Reference to the charge-conjugate decays is implied here and throughout the
  text.
\end{thebibliography}
%% \bibliographystyle{h-physrev3}
%\bibliographystyle{apsrev}

 % because the bibtex-generated one didn't do
                         % PRL-style formatting correctly
\end{document}